\documentclass[preprint,12pt,authoryear]{elsarticle}

\usepackage{amssymb}
\usepackage{xr-hyper}
\usepackage{amsmath}
\usepackage{booktabs}
\usepackage{graphicx}
\usepackage{tabularx}
\usepackage{makecell}
\usepackage{array}
\usepackage{threeparttable}
\usepackage{natbib}
\usepackage{multirow}
\usepackage{enumitem}
\journal{Computers, Environment and Urban Systems}
\usepackage{hyperref}
\usepackage{xr-hyper}
\begin{document}
\begin{frontmatter}



\title{CoRenew: A policy simulator based on large language model agents
for multifamily residential redevelopment}

\author[aff1,aff2]{Yudi Zhang}
\author[aff1]{Yuming Lin}
\author[aff1]{Li Tian\corref{cor1}}
\ead{litian262@126.com}
\author[aff1]{Yu Wang}
\author[aff3]{Jianghao Yu}

\cortext[cor1]{Corresponding author}

\affiliation[aff1]{
    organization={Department of Urban Planning, Tsinghua University},
    city={Beijing},
    postcode={100084},
    country={China}
}

\affiliation[aff2]{
    organization={Department of City and Regional Planning, UC Berkeley},
    city={Berkeley},
    state={CA},
    postcode={94720},
    country={USA}
}

\affiliation[aff3]{
    organization={Department of Electronic Engineering, Tsinghua University},
    city={Beijing},
    postcode={100084},
    country={China}
}

%% Abstract
\begin{abstract}
Evaluating policy effects in multifamily residential redevelopment is challenging because the process involves multi-round negotiations among stakeholders. Existing ex ante policy evaluation models struggle to capture such adaptive responses because they rely on predefined behavioral rules. Here, we present CoRenew, an open-source platform for evaluating policy effects by simulating multi-round negotiations with LLM-based agents. Using publicly available geographic and demographic data, CoRenew generates synthetic residents, selects weighted representatives, and compares policy alternatives across multiple objectives. We validate the platform against survey responses from 324 residents and a nine-month redevelopment negotiation observed in a real-world case. The simulated structural model recovered 19 survey-based paths, with 100\% agreement in directional signs and 94.7\% agreement in statistical significance. The best-performing LLM also closely matched the observed negotiation trajectory. Furthermore, we demonstrate CoRenew’s use in a Huadu District case study through three applications: global policy optimization, community-specific policy design, and the assessment of semantic policy interventions. The results show that CoRenew can reveal potential equity risks associated with consensus-promoting policies and suggest nonlinear subsidy effects: reductions in inequality become apparent only at higher subsidy levels, while utility gains for low-income residents exhibit diminishing returns. Our study demonstrates the value of incorporating negotiation dynamics into inclusive housing policy design and presents CoRenew as a transferable, visualization-enabled decision-support tool across different institutional contexts.
\end{abstract}

% %% Graphical abstract
% \centering
% \includegraphics[width=1\textwidth]{figures/00_graph_abstract.png}

% %% Research highlights

% \begin{highlights}
% \item An open-source LLM platform for numerical and semantic multifamily residential redevelopment policy assessment.
% \item Two-level adaptive framework linking global policies with local negotiation outcomes for policy assessment and refinement.
% \item Built-in visualization modules enable spatial outcome mapping, negotiation playback, and multi-objective policy comparison.
% \item CoRenew’s LLM agents recovered survey-based causal paths and the observed nine-month negotiation pathway.
% \item CoRenew revealed equity risks and nonlinear subsidy effects in public subsidy design.
% % \item LLM-based resident agents use theory-of-planned-behavior-guided reasoning prompts to enable context-sensitive negotiation.
% % \item Validation using survey responses from 324 residents and a nine-month traced negotiation process from a real redevelopment case.

% \end{highlights}

%% Keywords
\begin{keyword}
multifamily residential redevelopment \sep large language models \sep generative agent-based modeling \sep negotiation simulation \sep policy evaluation
\end{keyword}

\end{frontmatter}

\section{Introduction}
\label{sec1}

As the stock of multifamily residential buildings has expanded worldwide, their redevelopment has become increasingly important for housing supply, urban regeneration, and sustainable urban development in growing metropolitan areas \citep{lujanen2010legal,chia2023redevelopment}. The redevelopment of multifamily residential buildings often involves lengthy negotiations among developers, planners, and residents to obtain consent from existing owners and reach agreement on purchase prices or compensation arrangements \citep{cheng2019does}. The presence of multiple heterogeneous owners makes collective action even more difficult \citep{puustinen2015infill}, leading to holdout behavior, high transaction costs, and objections from minority owners \citep{lin2014strategic,Zho23}. The persistent gap between policy design and implementation outcomes leaves aging residential buildings unredeveloped, posing substantial safety risks \citep{Puu18}.

To address this implementation gap, a growing number of studies have developed simulation models to support the ex ante assessment of redevelopment policies \citep{hammond2015assessing}. Existing studies have used agent-based modeling (ABM) and equilibrium-based approaches to represent heterogeneous actors and examine bounded rationality, social networks, and the diffusion of information, norms, and behaviors \citep{xi2025rise,huang2023evolution,Wan25b,Yua24b,Zho23}. By incorporating actor heterogeneity, these models enable researchers to design more adaptive policies for different subpopulations and contexts \citep{savin2023agent}. Examples include adaptive subsidy mechanisms based on the proportion of cooperative developers or residents, as well as incentive strategies tailored to community social networks and initial willingness levels \citep{chu2020evolutionary,huang2023evolution}. These models also help policymakers compare the effects of different policy instruments through \textit{in silico} experimentation, thereby supporting more effective consensus-promoting policy design \citep{huang2025agent}.

Despite these advances, rule-based policy-evaluation models remain limited in their representation of both micro-level behavior and macro-level policy space. At the micro level, stakeholder decisions are shaped by the evolving dynamics of negotiation rather than by fixed incentives alone. Prior exchanges influence how subsequent proposals are interpreted, while the sequencing of offers can alter expectations, reference points, and willingness to compromise \citep{elfenbein2023emotion}. These context-dependent and path-dependent mechanisms are difficult to represent adequately through predefined behavioral rules \citep{boothby2023embracing}. At the macro level, ABMs and related approaches are typically constrained by numerical parameterization, requiring policies to be translated into explicit computational inputs and limiting assessment to a predefined set of scenarios \citep{lin2014strategic,huang2023evolution,huang2025agent}.  In practice, however, redevelopment policy assessment often involves flexible combinations of policy instruments and semantic institutional arrangements, such as escrow accounts and resident supervision mechanisms \citep{lujanen2010legal,Ho13,Cro20}. Existing models therefore have limited capacity to evaluate such policies or generalize across broader policy configurations \citep{zhang2026gplab}. 

Generative agent-based modeling (GABM) offers a promising approach to addressing these limitations by embedding large language model (LLM)-based agents within traditional agent-based models as cognitive engines for social interaction and decision-making \citep{ghaffarzadegan2024generative,xi2025rise}. Unlike agents governed by fixed response rules, LLM-based agents can generate context-sensitive decisions, maintain role-specific preferences, and adapt their positions over multi-round interactions through prompting, in-context learning, and fine-tuning \citep{gao2024large,wang2026negotiating,xi2025rise}. Recent studies have further demonstrated their capacity to produce coherent and adaptive behavior in deliberation, negotiation, and bargaining settings \citep{abdelnabi2024cooperation,wang2026negotiating}. In addition, the natural-language understanding capabilities of LLMs enable GABM to represent semantic policy inputs that are difficult to encode using numerical parameters alone \citep{zhang2026gplab}. General-purpose GABM frameworks have recently emerged in fields such as economic policy optimization and public health interventions \citep{liu2025agentic,zhang2026gplab}. However, the potential of GABM for urban policy design remains underexplored, and micro-level negotiation processes are often overlooked in general-purpose policy simulation frameworks. Its broader application is further constrained by privacy concerns surrounding individual-level profile data, the computational cost of large-scale simulations, and the scarcity of real-world negotiation data for empirical validation.

To address these gaps, we present CoRenew, a policy simulation platform for multifamily residential redevelopment that integrates negotiation simulation, policy assessment, and result visualization. Conceptually, CoRenew adopts a two-level framework linking individual stakeholder responses with policy settings through an iterative cycle of policy formulation, simulated stakeholder feedback, and policy refinement. Technically, the platform generates synthetic residents from publicly available datasets and employs a weighted representative-selection mechanism to reduce computational costs without disregarding population heterogeneity. Resident agents are modeled within a Markov game, with role-specific observation spaces, action spaces, and utility functions. Their reasoning is further structured through a chain-of-thought process grounded in the theory of planned behavior (TPB) \citep{ajzen1991theory}. To assess behavioral realism, we validate CoRenew against survey responses from 324 residents and a nine-month negotiation process observed in a real redevelopment case. Across 19 comparable structural paths, the simulated model achieved 100\% agreement in directional signs and 94.7\% agreement in statistical significance. Moreover, the best-performing LLM closely reproduced the observed negotiation trajectory. We further demonstrate CoRenew through a real-world application in Huadu District, Guangzhou, China. The application illustrates the platform’s capacity to identify globally optimal policies across competing objectives, derive community-specific policy designs, and assess semantic policy interventions.

\section{Related work}

\subsection{Residents' behavior in multifamily residential building redevelopment}
A substantial body of research has examined residents’ behavior in multifamily residential redevelopment by focusing on both psychological motivations and external socio-environmental factors \citep{Yam17b,Gao16b}. Behavioral theories help identify the psychological factors that shape residents’ willingness to participate in redevelopment, including the theory of planned behavior, cumulative prospect theory, and social cognitive theory \citep{stajkovic1998social,ebrahimigharehbaghi2022application,tang2022residents}. Among these frameworks, the theory of planned behavior (TPB) is one of the most commonly used \citep{huang2023evolution}. TPB posits that behavioral intention is shaped by three key constructs: attitude, subjective norms, and perceived behavioral control \citep{ajzen1991theory}. Attitudes capture how owners evaluate redevelopment in terms of expected costs, benefits, risks, and distributive fairness. Subjective norms reflect perceived expectations, social pressure, and interpersonal influence within residential networks. Perceived behavioral control refers to whether owners view redevelopment as practically achievable under existing organizational, financial, and institutional conditions, including whether they believe collective coordination is likely to succeed. In this study, we use the three TPB constructs to structure resident-agent reasoning and validate the resulting causal structures by comparing them with those estimated from resident survey data.

Resident decision-making is also influenced by external socio-environmental factors, such as group size, resident heterogeneity, building condition, and location \citep{cheng2019does,huang2025agent,gao2016does,Ho13}. Larger groups and greater heterogeneity among residents increase coordination costs and make collective consent more difficult to achieve. Empirical studies have also examined spatial and environmental factors, showing that building condition shapes perceived urgency, while building location affects redevelopment profitability and expected gains. However, high-value locations may also intensify bargaining over surplus distribution \citep{gao2016does}. These psychological and socio-environmental factors interact and jointly shape residents’ redevelopment decisions \citep{Ti23}.

\subsection{Policy design for multifamily residential redevelopment}

As multiple stakeholders are involved in urban renewal negotiations, the literature has examined various institutional strategies for facilitating collective consent and empowering residents. One major strand focuses on financial instruments, such as reserve funds, collective loans, subsidies, tax incentives, and value capture through infill development or additional development rights \citep{puustinen2015infill,Puu17}. These instruments primarily operate by reducing immediate private payment burdens and increasing expected redevelopment gains. However, they may also generate inequitable compensation schemes and uneven distributions of benefits, potentially benefiting mainly middle-income residents \citep{puustinen2015infill,Puu17,Puu18}.

Beyond financial tools, another strand highlights procedural and institutional arrangements that address distrust, information asymmetry, and other potential barriers to negotiation \citep{lujanen2010legal,Ho13,Cro20}. Common instruments include innovative benefit-sharing arrangements \citep{Puu17,Puu18}, building management associations \citep{Men03}, majority voting rules \citep{Ti23}, termination procedures, and measures to enhance procedural transparency \citep{easthope2013urban,Ti23}. The positive role of institutional interventions in urban renewal has been well documented, as such interventions can empower residents and indirectly shape negotiation outcomes \citep{Cro20}. However, existing studies are mainly based on institutional comparisons or case studies and rarely examine how financial incentives and procedural arrangements reshape micro-level negotiation dynamics. As a result, their policy recommendations tend to be context-specific and offer limited capacity for ex ante assessment of collective consent outcomes across different redevelopment contexts.

\subsection{Modeling approaches for housing policy assessment}

Policy simulation models provide a means of examining policy processes across interconnected dimensions of population, institutional design, behavior, and time \citep{schunemann2024complex}. By representing the adaptive responses of heterogeneous stakeholders, modeling approaches enable decision-makers to anticipate how alternative policy configurations may shape collective agreement and generate unintended consequences prior to implementation \citep{cheng2019does}. As summarized in Table \ref{tab:model_comparison}, existing approaches to housing policy assessment can be broadly organized into three strands: evolutionary game-theoretic models, agent-based models, and, more recently, learning-based adaptive systems.

Evolutionary game theory is primarily concerned with the emergence of stable strategies in evolving populations and the evolution of strategic behavior under dynamic conditions \citep{weibull1997evolutionary}. In residential redevelopment research, it has been widely used to represent stakeholders’ strategic choices between cooperation and non-cooperation, identify equilibrium outcomes, and derive context-specific policy implications for facilitating collective action \citep{Wan25b,Yua24b,Zho23}. For example, \citet{chu2020evolutionary} developed a three-population evolutionary game model and found that adaptive subsidies were more effective than fixed subsidies in reducing strategic fluctuations and sustaining cooperation. Similarly, \citet{melo2017effect} examined owners’ investment decisions in urban housing renewal and found that positive externalities generated by surrounding rehabilitation may encourage wait-and-see behavior, thereby discouraging private actors from investing. However, game-theoretic models often represent residents as a relatively homogeneous population and devote limited attention to within-group heterogeneity and its influence on collective outcomes.

ABM offers a complementary approach by representing interactions among heterogeneous individuals, as well as between individuals and their environments, through explicit behavioral rules \citep{furtado2019policy,huang2025agent}. Recent applications to multifamily residential redevelopment have examined how community social networks, opinion dynamics, and heterogeneous willingness to participate influence policy outcomes \citep{wang2016simulation,huang2023automatic,huang2025agent}. A further strand comprises learning-based methods. Although these approaches have not yet been directly applied to housing policy, they have enabled the development of two-level policy optimization frameworks in economic domains such as taxation and pension design \citep{zheng2022ai,mi2025econgym}. Despite these advances, existing policy simulation frameworks remain subject to several limitations. Their reliance on predefined behavioral rules tends to overlook context-dependent negotiation dynamics, while their representation of policy is often restricted to explicit numerical parameters or predefined scenarios \citep{gao2024large,zhang2026gplab}. Moreover, parameterizing the reciprocal relationship between policy design and negotiation outcomes can substantially increase both computational demands and model complexity \citep{schunemann2024complex}. This creates a persistent trade-off between behavioral richness and policy generalizability: models capable of representing detailed negotiation processes are often difficult to scale, whereas those designed for broad policy comparison typically rely on simplified assumptions about stakeholder behavior.

More recently, GABM has emerged as a promising approach for representing micro-level social dynamics and assessing the behavioral consequences of policy interventions \citep{huang2026policysim,gao2024large}. By employing generative models such as LLMs as cognitive engines, GABM seeks to represent human decision-making in interactive social settings \citep{wang2026negotiating,liu2025agentic}. Although many proposed frameworks remain conceptual, several implemented platforms have begun to demonstrate the potential of GABM for social and policy experimentation. AgentSociety provides an open-source, large-scale social simulation platform that integrates LLM-driven agents with urban, social, and economic environments and a distributed simulation engine \citep{piao2025agentsociety}. It has been used to simulate large populations and examine social and policy interventions, including universal basic income and responses to external shocks, with an emphasis on emergent social behavior and scalable computational experiments. GPLab more directly operationalizes a general-purpose policy laboratory through three interconnected components: LLM-based social agents, modular social subsystems, and a simulation and evaluation engine \citep{zhang2026gplab}. The framework simulates heterogeneous individual responses to consumption-voucher and carbon policies, focusing particularly on policy transmission across economic and public-opinion subsystems. However, existing GABM platforms have devoted limited attention to micro-level stakeholder negotiation and spatially differentiated urban contexts, constraining their direct application to urban policy evaluation.

\begin{table*}[htbp]
\centering
\caption{Comparison of simulation frameworks for housing redevelopment and broader policy optimization}
\label{tab:model_comparison}

\small
\renewcommand{\arraystretch}{1.08}
\setlength{\tabcolsep}{4pt}

\begin{minipage}{\textwidth}
\centering

\resizebox{\textwidth}{!}{%
\begin{tabular}{p{3.2cm} p{2.8cm} p{2.2cm} p{1.8cm} p{1.8cm} p{1.9cm} p{3.0cm} p{2.7cm}}
\toprule
\textbf{Study} & \textbf{Application} & \textbf{Policy input} & \textbf{Executable system} & \textbf{Individual profiling} & \textbf{Validation}& \textbf{Agent modeling approach} & \textbf{Policy optimization} \\
\midrule

\citet{Lia19}
& Energy-efficiency retrofit
& Numerical& —& —& —& Rule-based ABM
& —\\

\citet{huang2023evolution,huang2025agent}
& Community redevelopment& Numerical& —& Yes
& —& Rule-based ABM
& —\\

\citet{Wan25b,Yua24b}
& Community redevelopment& Numerical& —& —& Single case& Evolutionary game
& —\\

\citet{Zho23}
& Community redevelopment& Numerical& —& —& Single case& Evolutionary game + Deep RL
& —\\

\citet{Bad24}
& Affordable housing provision and planning
& Numerical& —& —& Trend & Rule-based ABM
& —\\

\citet{Gao24}& Participatory design
& —& —& Yes
& —& LLM-based agents
& —\\

\citet{abdelnabi2024cooperation}& Negotiation& —& Yes& Yes& —& LLM-based agents&—\\

\citet{wang2026negotiating}& Community redevelopment& —& —& Yes
& Single case& LLM-based agents
& —\\

\citet{zheng2022ai}
& Taxation policy optimization
& Numerical& Yes
& Yes
& Distribution& Rule-based ABM
& RL \\

\citet{mi2025econgym}
& Economic policy optimization
& Numerical& Yes
& Yes& Distribution& Rule-based ABM
& RL / LLM / Rule-based \\

\midrule
\textbf{CoRenew}
& \textbf{Community redevelopment}& \textbf{Numeric and semantic policy}& \textbf{Yes}
& \textbf{Yes}
& \textbf{Mechanism \& single case \& trend}& \textbf{LLM-based agents}
& \textbf{LLM / Rule-based}\\

\bottomrule
\end{tabular}%
}

\end{minipage}
\end{table*}

\section{CoRenew}
\begin{figure*}[htbp]
    \centering
    \includegraphics[width=0.8\textwidth]{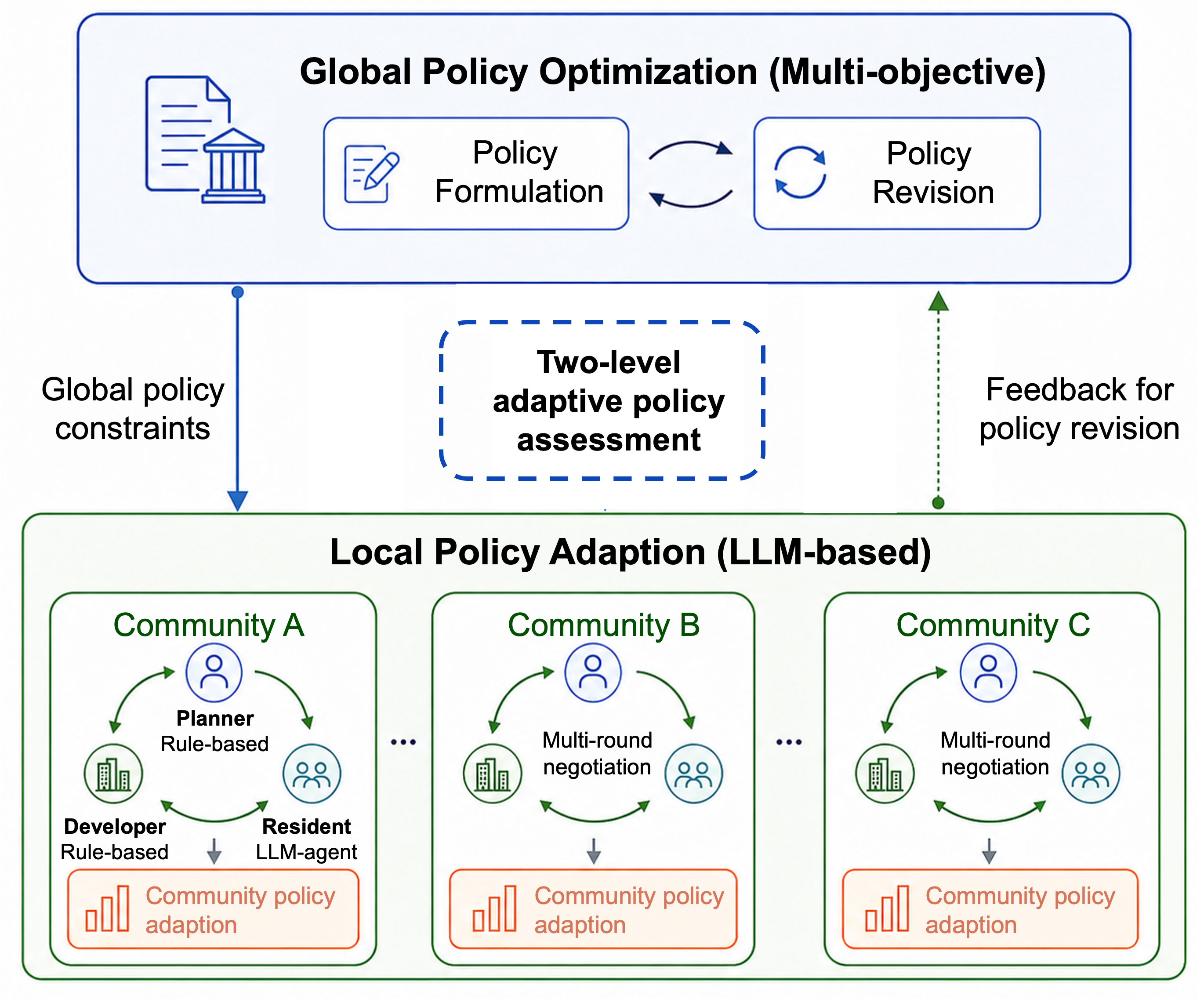}
    \caption{Two-level adaptive framework linking global policy design with local negotiation outcomes for policy assessment and refinement.}
    \label{fig:conceptual_framework}
\end{figure*}

CoRenew is an open-source platform for simulating negotiations in multifamily residential redevelopment and supporting the evaluation and refinement of redevelopment policies. The platform follows a two-level iterative policy framework (Figure \ref{fig:conceptual_framework}). At the macro level, users define the policy parameters, constraints, and evaluation objectives to be tested. At the micro level, the system simulates multi-round negotiations within each community to examine how global policies are adapted and implemented under local conditions. This process reflects recurring negotiation mechanisms documented in multifamily residential redevelopment, where planners, developers, and residents repeatedly adjust their decisions in response to evolving proposals and stakeholder reactions \citep{wang2026negotiating,muczynski2023collective}.

Within each community, planner, developer, and resident agents participate in multiple rounds of negotiation while pursuing distinct objectives. The planner proposes community-specific public interventions within the global policy constraints, while the developer formulates a financially feasible redevelopment proposal. Resident agents evaluate the proposed policy and redevelopment package, decide whether to support or oppose redevelopment, and specify the conditions under which they would agree or make counteroffers. Based on residents’ responses, the planner and developer independently adjust their respective policy instruments and project proposals. This process continues until an agreement is reached or predefined termination conditions are met. The resulting locally adapted policies and negotiation outcomes are then fed back to the macro level and evaluated against multiple objectives.

As a simulation tool, CoRenew allows users to integrate real-world geographic and population data, construct heterogeneous resident, planner, and developer agents, and simulate negotiations under different policy and contextual conditions. As a policy-evaluation tool, it supports a wide range of numerical and semantic policy inputs and flexible objective settings. The platform visualizes both aggregate policy performance and community-specific outcomes, enabling public-sector users to identify where a policy succeeds, fails, or produces uneven effects and to refine the policy accordingly.

\begin{figure*}[htbp]
    \centering
    \includegraphics[width=0.9\textwidth]{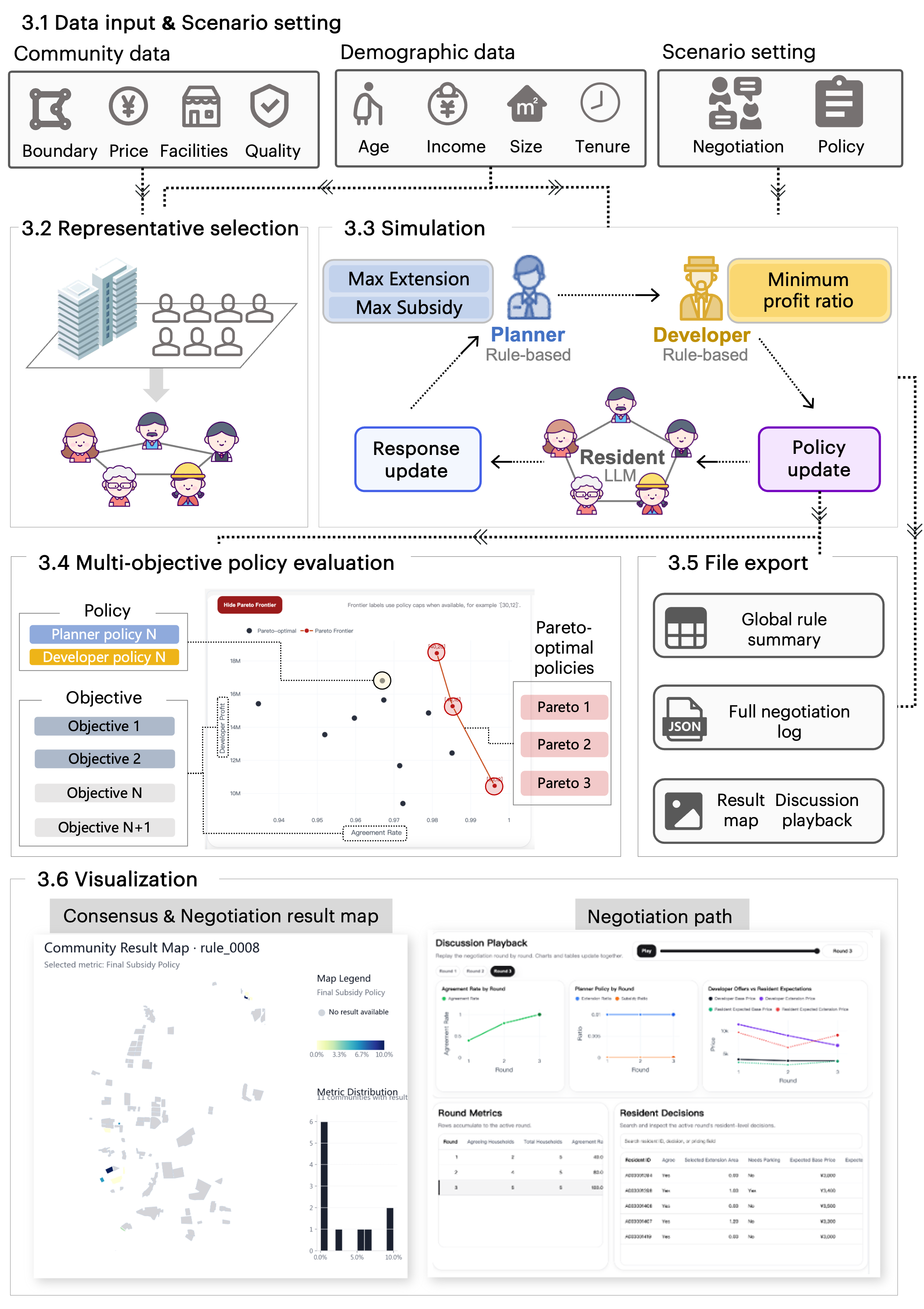}
    \caption{Overall framework}
    \label{fig:overall_framework}
\end{figure*}

\clearpage
The CoRenew workflow consists of six sequential stages (Figure \ref{fig:overall_framework}). In (1) data input and scenario setting, users provide the empirical inputs required to initialize the model and define the institutional and procedural conditions of negotiation. In (2) representative selection, CoRenew generates synthetic resident profiles from the uploaded data and selects representative agents for each community. In (3) simulation, CoRenew simulates multi-round negotiations among stakeholders over housing redevelopment under the specified policy constraints. In (4) multi-objective evaluation, the simulation outputs are assessed under alternative objective settings, allowing users to compare multiple policy scenarios and identify Pareto-optimal policy combinations either overall or for specific communities. In (5) file export, users can export the global rule summary, full negotiation log, result maps, and discussion playback records in multiple file formats compatible with external software. In (6) visualization, the platform presents both consensus and negotiation outcome maps, as well as the negotiation path, enabling users to inspect the spatial distribution of outcomes and the round-by-round dynamics of stakeholder interactions.

\subsection{Data input and scenario setting}

CoRenew is initialized with two categories of inputs: the community information table and scenario settings, which are summarized in Appendix Table~\ref{tab:model_inputs}. The community information table contains the core attributes required for simulation, including project boundaries, facilities accessible within a 15-minute walk, housing prices, parking conditions, building quality, relevant governance arrangements, and community-level demographic variables such as dwelling size and tenure structure. To facilitate implementation, CoRenew provides a downloadable template file that includes all required variables. 

Scenario settings define the policy, utility, and negotiation settings in the simulation. Policy settings specify the parameter ranges for planners (e.g., maximum subsidy ratio) and developers (e.g., minimum profit rate). CoRenew also accepts semantic policy inputs, allowing non-financial instruments to be incorporated into the simulation. Utility settings represent stakeholder objectives through user-configurable modules. Users should define the composition of costs and benefits for each type of agent. An example default utility specification is provided in Appendix~\ref{appendix_utility}. Negotiation settings specify the discussion protocol, including the number of discussion rounds and the termination conditions.

\subsection{Synthetic population construction and representative selection} \label{profile_generation}

Because individual-level microdata are often unavailable or restricted by privacy concerns, CoRenew includes a resident-generation module that constructs virtual agents from publicly available census-unit data. The inputs include household size, vacancy rates, income and gender distributions, and hardship rates. The module also draws on community-level data, such as housing prices, predominant dwelling sizes, and resident counts, to generate community-specific agents. The generation procedure sequentially reconstructs the housing-unit structure, assigns tenure categories, maps households to the local income distribution, determines household size, and finally samples individual-level attributes. The resulting synthetic population is designed to match the observed socioeconomic composition of the community while preserving within-community heterogeneity. Detailed generation rules and parameter settings are provided in Appendix~\ref{appendix_profile_generation}.

CoRenew further incorporates a representative-selection module to reduce simulation costs while preserving community heterogeneity and ensuring the representation of socioeconomically disadvantaged residents. The module is motivated by the assumption that residents are more likely to trust and be represented by individuals with similar socioeconomic profiles \citep{mansbridge2009selection}. Representative selection therefore considers both the statistical representativeness of individual residents and the risk that disadvantaged groups may be overlooked when selection is driven primarily by dominant community profiles.

Each resident is embedded in a multidimensional feature space defined by age, income, housing area, household size, tenure status, and housing type. For each community \(c\), pairwise distances are converted into representative-selection probabilities using an exponential decay kernel:
\begin{equation}
p_{ijc}
=
\frac{\exp(-d_{ijc}/\tau_c)}
{\sum_{k=1}^{N_c}\exp(-d_{ikc}/\tau_c)},
\end{equation}
where \(d_{ijc}\) denotes the distance between residents \(i\) and \(j\), \(N_c\) is the number of residents in community \(c\), and \(\tau_c\) is a community-specific distance-decay parameter defined as the median of all non-zero pairwise distances within that community. The representative potential of resident \(j\) is calculated as:
\begin{equation}
w_{jc}^{*}
=
\sum_{i=1}^{N_c}p_{ijc}.
\end{equation}

To prevent dominant socioeconomic groups from disproportionately shaping the selected representatives, residents are partitioned into \(K\) clusters in the same feature space using K-means, where \(K\) is the user-defined number of representatives per community. Within each cluster, the resident with the highest representative potential is selected. The module also reserves at least one representative seat for low-income residents, defined as those with annual incomes below the median income of the target area. If the initial selection contains no low-income resident, the selected representative with the lowest representative potential is replaced by the low-income resident with the highest representative potential. This safeguard reflects evidence that collective consent in multifamily housing redevelopment may be delayed or prevented when residents facing financial constraints or personal difficulties cannot accept the proposed scheme \citep{Yam17b}.

After the final representative set \(\mathcal{R}_c\) has been determined, each resident is assigned to the representative with the most similar profile. The voting weight of representative \(r\) is defined as the number of residents assigned to that representative:
\begin{equation}
\omega_{rc}
=
\sum_{i=1}^{N_c}
\mathbb{I}
\left(
r=\arg\min_{k\in\mathcal{R}_c}d_{ikc}
\right).
\end{equation}
Thus, \(\sum_{r\in\mathcal{R}_c}\omega_{rc}=N_c\). When the number of residents does not exceed \(K\), all residents are retained as representatives and assigned unit weights.

Each representative's agreement decision is propagated to the residents assigned to that representative. The community agreement rate is therefore calculated as:
\begin{equation}
A_c
=
\frac{1}{N_c}
\sum_{r\in\mathcal{R}_c}
\omega_{rc}y_{rc},
\end{equation}
where \(y_{rc}\in\{0,1\}\) indicates whether representative \(r\) supports the final redevelopment proposal. Representatives are used only to approximate residents' agreement decisions. Utility is calculated separately for every synthetic resident using the resident's own socioeconomic and housing attributes and the applicable policy and redevelopment conditions. Community-level welfare and inequality measures are then calculated directly from these individual-level utilities.

\subsection{Simulation}
\subsubsection{Agent framework}
CoRenew is built around three core roles in residential redevelopment: residents, developers, and planners. Each role is represented by a distinct agent type in a Markov game, with role-specific observation spaces, action spaces, and decision policies. Given private observations, agents select actions according to their policies, after which the environment updates to reflect changes in policy settings, agreement status, and negotiation conditions. In each negotiation round, agents observe current project conditions, institutional constraints, and other participants’ actions, and then take role-specific actions that jointly determine the evolution of the negotiation state. This design allows heterogeneous agents to interact under diverse redevelopment scenarios. Residents are modeled as LLM-based agents to capture heterogeneity in resident profiles and the complexity of negotiation dynamics. Planners and developers, by contrast, are implemented using rule-based algorithms that adapt their strategies to resident responses, improving overall system stability. Table~\ref{tab:agent_setup} summarizes the observations and actions for all agent types in CoRenew. 

\begin{table*}[htbp]
\centering
\caption{Observations and actions of agent types in CoRenew}
\label{tab:agent_setup}

\small
\renewcommand{\arraystretch}{1.15}
\setlength{\tabcolsep}{4pt}

\begin{tabular}{@{}>{\raggedright\arraybackslash}p{1.8cm}
                >{\raggedright\arraybackslash}p{2.8cm}
                >{\raggedright\arraybackslash}p{2.6cm}
                >{\raggedright\arraybackslash}p{3.0cm}
                >{\raggedright\arraybackslash}p{3.0cm}@{}}
\toprule
\textbf{Agent type} & \multicolumn{2}{c}{\textbf{Observation}} & \multicolumn{2}{c}{\textbf{Action}} \\
\cmidrule(lr){2-3} \cmidrule(lr){4-5}
& \textbf{Public} & \textbf{Private} & \textbf{Public} & \textbf{Private} \\
\midrule

Resident
&
\multirow{3}{2.8cm}{\raggedright
agreement rate; \newline
approval threshold; \newline
rounds left; \newline
policy limits
}
&
cost and utility; \newline
housing condition
&
support or oppose; \newline
quoted base price; \newline
quoted extension price; \newline
stated parking decision; \newline
quoted extension area
&
reservation base price; \newline
reservation extension price; \newline
parking willingness to pay; \newline
desired extension area
\\

\cmidrule(lr){1-1}\cmidrule(lr){2-5}

Developer
&
same public observations as residents&
project cost and revenue; \newline
minimum profit requirement
&
set base price; \newline
set extension price; \newline
set parking price
&
---
\\

\cmidrule(lr){1-1}\cmidrule(lr){2-5}

Planner
&
same public observations as residents&
subsidy capacity; \newline
extension constraints; \newline
recent offer change
&
adjust subsidy; \newline
adjust extension ratio; \newline
choose policy package
&
---
\\

\bottomrule
\end{tabular}
\end{table*}

\subsubsection{Resident agent}
Resident agents are heterogeneous actors who decide whether to support redevelopment, whether to purchase parking spaces, and how much additional floor area to acquire. These decisions are shaped by both household-specific economic incentives and boundedly rational influences, such as conformity to collective opinion. Their responses in turn affect developer profitability and planner intervention. 

To make resident behavior more realistic, we incorporated two mechanisms. First, each agent follows a structured chain-of-thought procedure in a hidden scratchpad, which allows it to organize intermediate reasoning and private calculations \citep{abdelnabi2024cooperation}. In each round, the resident observes its current cost and utility plan, the negotiation state, and the relevant policy constraints. It then formulates a private reservation offer and a public counteroffer concerning the redevelopment cost. After considering all these observations, it decides whether to accept the proposal. To improve stability, proposals meeting the private acceptance threshold are automatically accepted.

Second, resident prompting is structured using the Theory of Planned Behavior. Decision-relevant information is organized around three dimensions: \emph{attitude}, \emph{perceived behavioral control}, and \emph{subjective norm}. \emph{Attitude} is represented by the expected benefits of renewal and current housing conditions, \emph{perceived behavioral control} by the financial burden of participation, and \emph{subjective norm} by the visible negotiation state, including the current agreement rate and the approval threshold. The full prompt template is shown in Appendix \ref{app:resident_prompt}.

\subsubsection{Planner agent}
Planner agents are institutional actors responsible for policy intervention during negotiation. Their actions are constrained by external planning regulations, including the maximum admissible extension and subsidy ratios. The planner follows the principle of minimum necessary intervention. It adjusts policy only when developer profitability turns negative, the negotiation reaches its midpoint without meeting the required approval threshold, or developer offers remain nearly unchanged for two consecutive rounds. Once an intervention is triggered, the planner samples candidate increases in subsidy and extension ratio from the remaining feasible policy space and selects the policy most likely to induce additional resident support with the minimum intervention.

\subsubsection{Developer agent}
Developer agents represent profit-constrained redevelopment firms that negotiate pricing terms with residents. As such redevelopment projects are difficult to implement and often face high coordination costs, developers are assumed to be willing to make concessions as long as the project remains financially viable. In each round, the developer generates candidate pricing packages based on the prices quoted by residents who currently oppose the proposal and retains only those satisfying a minimum profit-rate constraint. Among the feasible packages, it then selects the option that maximizes the share of supporting residents.

\subsubsection{Negotiation process}
Negotiation is modeled as a finite-round sequential process. In each round, the state captures the current agreement rate, the planner’s policy settings, the developer’s pricing and profitability conditions, and residents’ responses from the previous round. Agents move in a fixed order. The planner first proposes a policy package, including the extension ratio and subsidy ratio. The developer then responds with a pricing package. Residents subsequently make individual decisions. At the end of each round, the environment updates the aggregate outcomes, including the weighted community agreement rate, developer profit, profit margin, total extension area, and parking demand ratio. The process continues until either the required community approval threshold is reached or the maximum number of rounds is exhausted. If the threshold is met, the current policy-pricing package is treated as the final redevelopment agreement. Otherwise, the negotiation ends without agreement.

\subsubsection{Efficiency}
Table~\ref{tab:efficiency_stats} reports the computational cost of CoRenew implemented with DeepSeek-V3.2, measured in terms of response time and token usage at both the resident-agent and round levels.

\begin{table}[htbp]
\centering
\caption{Response time and token usage per resident agent and per round}
\label{tab:efficiency_stats}
\small
\begin{tabular}{llccc}
\toprule
Level & Metric & Mean & Min & Max \\
\midrule
Per resident& Response time (s) & 89.37 & 26.01 & 567.62 \\
& Prompt tokens & 2343.88 & 1712 & 3549 \\
& Completion tokens & 772.35 & 587 & 1008 \\
& Total tokens & 3116.23 & 2363 & 4331 \\
\midrule
Per round & Response time (s) & 486.07 & 234.55 & 1295.73 \\
& Prompt tokens & 11926.15 & 10443 & 13393 \\
& Completion tokens & 3830.01 & 3217 & 4295 \\
& Total tokens & 15756.16 & 14360 & 17236 \\
\bottomrule
\end{tabular}
\end{table}

\subsection{Policy evaluation}
\label{sec:policy_optimization}
To identify the optimal policy, CoRenew applies a multi-objective assessment that balances efficiency and equity. Specifically, the objectives include: (1) the number of communities reaching the required approval threshold, (2) total subsidy expenditure, (3) the average utility of low-income residents, (4) utility inequality among residents, and (5) the average developer profit rate.

\subsection{File export}
CoRenew provides an export module for downstream analysis, visualization, and reporting. Available outputs include geospatial result files, policy comparison tables, complete negotiation logs and visualization figures. Community-level result maps are exported in GeoPackage (GPKG) format, storing both spatial geometry and key indicators such as the final agreement rate, realized extension ratio, and realized subsidy ratio. Policy comparison results are exported as CSV files, including the metrics used in pairwise trade-off analysis and multi-objective comparison. Complete negotiation records are exported in JSON Lines (JSONL) format, preserving the round-by-round bargaining process for each community. Visual outputs, including result maps, negotiation plots, and policy trade-off figures, can be exported as PNG files, with SVG available for publication-quality graphics.

\subsection{Visualization}
\begin{figure*}[htbp]
    \centering
    \includegraphics[width=1\textwidth]{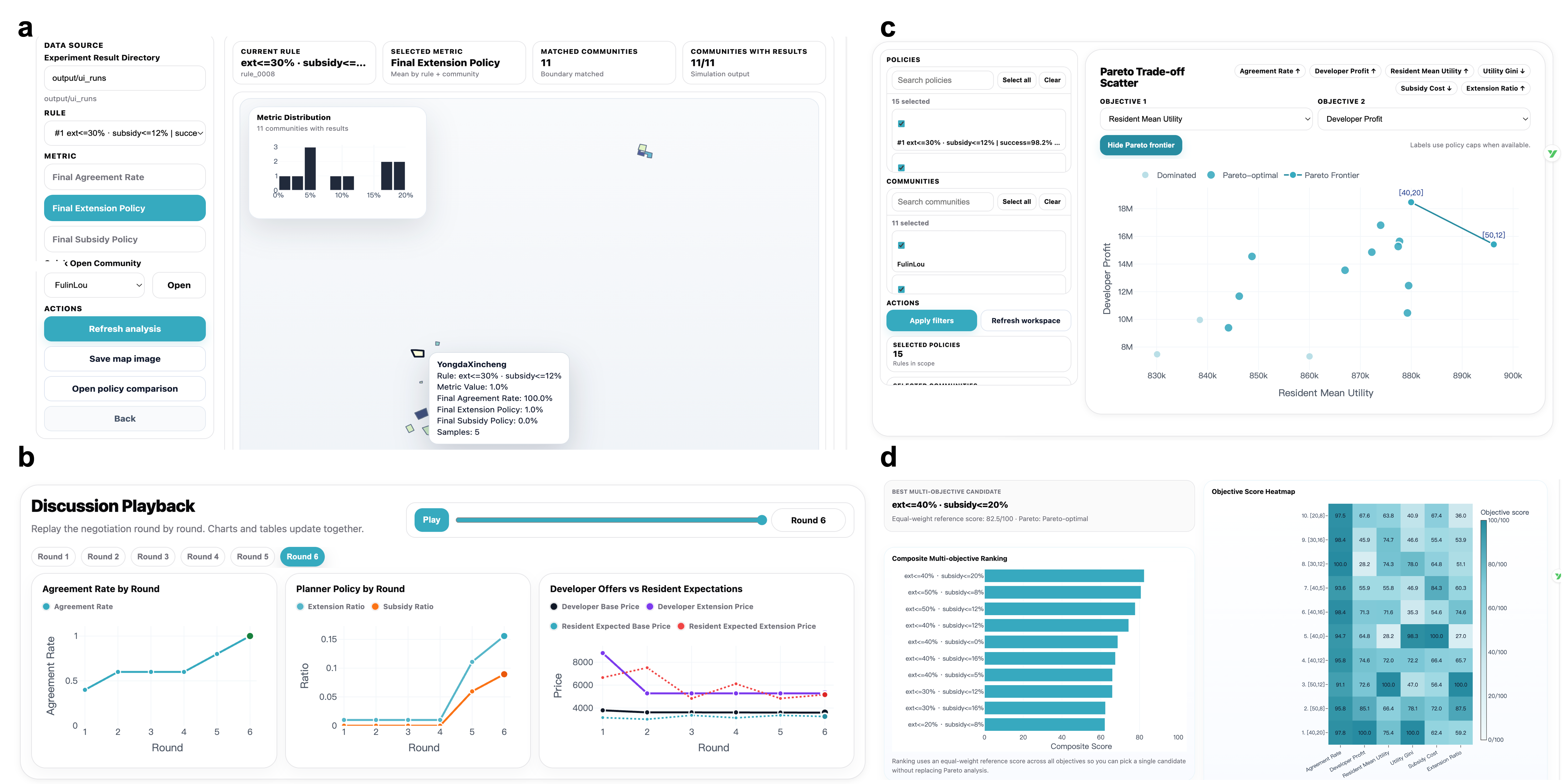}
    \caption{Visualization interface of CoRenew.}
    \label{fig:visualization}
\end{figure*}

CoRenew includes built-in visualization tools for inspecting negotiation outcomes and policy trade-offs. The interface comprises three modules: a community result map, a community-level negotiation viewer, and a policy trade-off analysis workspace. The community result map displays the negotiation outcome of each community under the selected policy (Figure \ref{fig:visualization}a). Users can visualize the final agreement rate, the planner's final extension ratio or the planner's final subsidy ratio. This enables direct comparison of outcome patterns and policy use across the study area. Each community can then be opened for detailed inspection (Figure \ref{fig:visualization}b). The community-level viewer summarizes the final outcome of the selected simulation and provides round-by-round playback of the negotiation process. Users can examine the evolution of agreement, planner adjustments, developer offers and resident expectations, together with round-level summary metrics and resident-level decisions. The policy trade-off analysis workspace supports comparison of multiple candidate policies under two objectives (Figure \ref{fig:visualization}c) as well as multiple goals (Figure \ref{fig:visualization}d). Users can choose the policies, communities, and evaluation objectives to include. For two selected objectives, the system visualizes policy performance in a two-dimensional trade-off space and identifies the Pareto-optimal set. For higher-dimensional evaluation, it provides a composite comparison view, including overall scores, rankings, objective profiles, and a detailed comparison table.

\subsection{Model validation}

We validated the model against both a real-world case and survey data. First, we examined whether the model could reproduce the observed negotiation process in a real-world redevelopment case. Details of the benchmark case are provided in Appendix \ref{huaducase}. We compared five LLMs, namely DeepSeek-V3.2, GLM, GPT-4.1, GPT-4o, and Gemini, and repeated each simulation 20 times to account for stochastic variation. Performance was evaluated at three levels: final outcome, negotiation process, and individual behavior. 

\begin{figure}[htbp]
    \centering
    \includegraphics[width=1.0\textwidth]{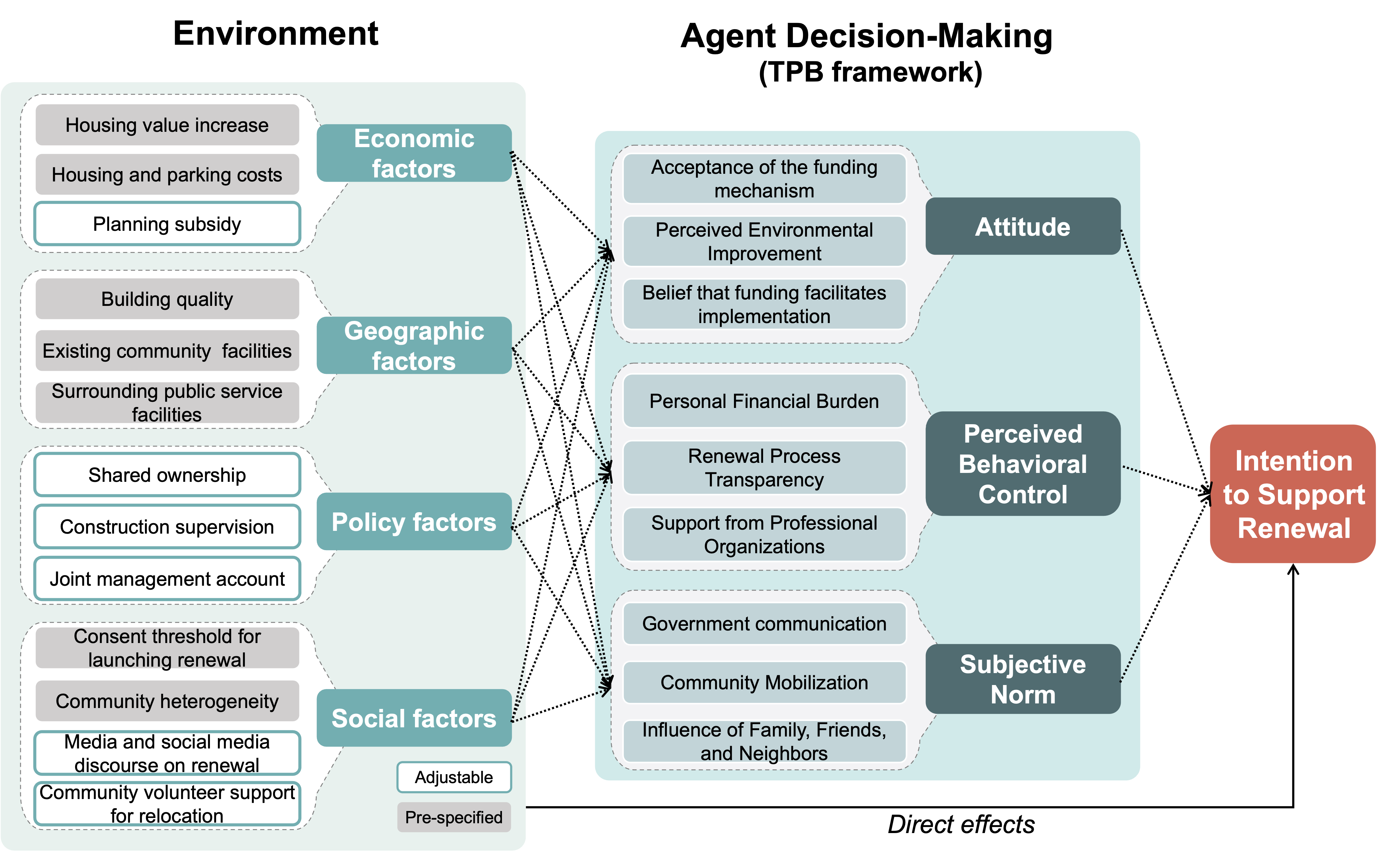}
    \caption{Structural model for comparing decision pathways between LLM agents and human survey respondents.}
    \label{fig:TPB_diagram}
\end{figure}

Second, we tested whether the simulated agents could reproduce the effects of economic, geographic, policy, and social factors on residents’ decisions by comparing structural equation path coefficients estimated from LLM-generated data with those estimated from real survey data. Figure~\ref{fig:TPB_diagram} presents the full structural equation model. We surveyed 324 residents about their willingness to participate in redevelopment and used questionnaire items to measure relevant environmental factors and specific TPB constructs. The resulting survey-based structural equation model was used as the benchmark. To estimate the corresponding paths for the LLM-based agents, we coded residents’ chain-of-thought reasoning using a predefined coding template implemented with DeepSeek-V3.2 (see Appendix~\ref{app:coding_template} for the full template). For each item, the coding output recorded its presence, directional valence, intensity, supporting textual evidence, and a brief explanation, together with quality checks for information sufficiency and contradiction detection. The coding results were further manually reviewed and validated to ensure accuracy. We then evaluated agreement with the survey benchmark in terms of coefficient magnitude, directional sign, statistical significance, and formal coefficient differences. Coefficient differences were assessed using a two-sided standardized difference test that accounts for uncertainty in both the survey-based and LLM-estimated coefficients. Appendix Table~\ref{tab:evaluation_metrics} summarizes all evaluation metrics used in the study.

\clearpage

\section{Results}
\subsection{Validation results}
\subsubsection{Validation against observed negotiation dynamics}
\begin{figure*}[htbp]
    \centering
    \includegraphics[width=0.8\textwidth]{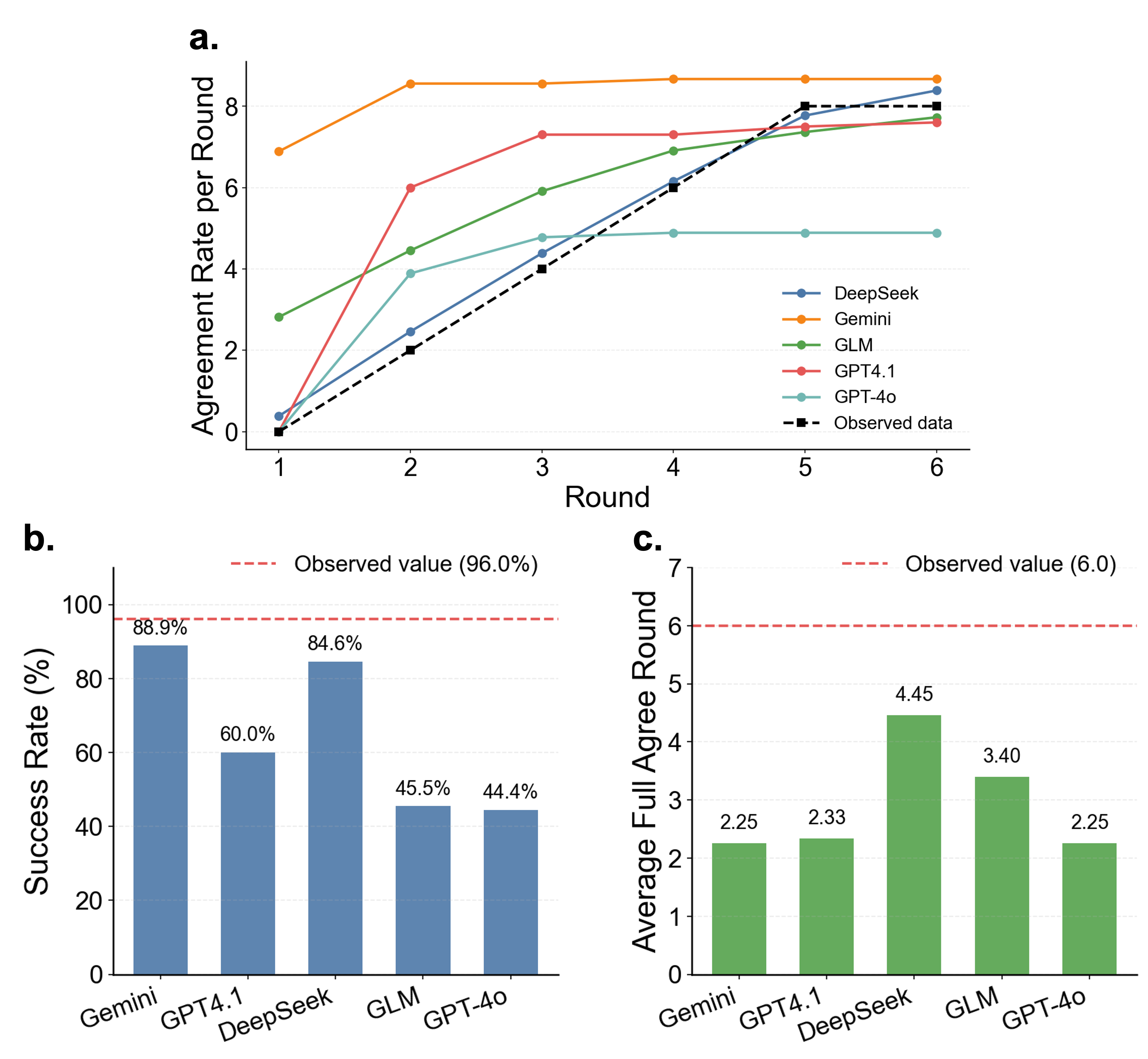}
    \caption{Comparison of LLM-simulated and observed consensus-building outcomes in the Huadu case under different models. (a) Round-by-round negotiation dynamics, measured by the cumulative agreement count during the discussion process. (b) Discussion success rate, defined as the proportion of simulation runs that reached the approval threshold. (c) Average round required to reach full agreement among successful runs.}
    \label{fig:huadu_validation}
\end{figure*}

Model performance varied substantially across LLMs, with DeepSeek-V3.2 showing the strongest overall correspondence to the observed negotiation process. For process-level accuracy, DeepSeek-V3.2 most closely reproduced the empirical negotiation pathway, yielding the lowest DTW distance (2.00; Figure \ref{fig:huadu_validation}a). Its agreement trajectory rose cautiously in the early rounds, accelerated only gradually through the middle stage, and stabilized later in the discussion, closely matching the observed pattern of delayed convergence. GLM also showed relatively strong process fidelity (DTW = 6.00): although it deviated from the empirical pattern in the early rounds, it aligned well with the observed trajectory in the later stage. By contrast, Gemini (DTW = 18.67), GPT-4.1 (DTW = 6.30), and GPT-4o (DTW = 10.00) generally underestimated the difficulty of negotiation. These models tended to generate high agreement rates too early, producing trajectories that converged prematurely and failed to capture the delays introduced by disagreement, repeated bargaining, and trade-off considerations that characterize real-world negotiation.

For final outcomes, LLM simulations generally produced both lower success rates and shorter paths to consensus relative to the observed case. In the real negotiation, the success rate was 96.0\%, and full agreement was reached on average in round 6. However, the highest simulated success rate was 88.9\% for Gemini, followed by 84.6\% for DeepSeek-V3.2, while all remaining models performed substantially worse. Moreover, among simulations that did reach unanimity, consensus was achieved markedly earlier than in the empirical case: the average round of full agreement ranged from 2.25 to 4.45 across models, well below the observed value of 6.0. The lower simulated agreement rates can be explained by the fact that, in actual negotiations, developers and planners often deploy additional interventions over time to increase support, such as joint management accounts and repeated informal communication \citep{wang2026negotiating,huang2025agent}. 

\subsubsection{Validation of TPB-based structural path estimates}
Based on the real-case validation, we selected DeepSeek-V3.2 for the multiple-project experiments because it achieved the strongest overall performance. The path estimates derived from DeepSeek-V3.2-generated negotiation data were broadly consistent with those from the real questionnaire data, especially in directional signs and statistical significance (Table~\ref{tab:real_llm_comparison_grouped}). Across 19 comparable paths, the LLM-based SEM recovered all benchmark paths, achieved 100\% accuracy in directional signs, and reached 94.7\% accuracy in statistical significance. The only significance mismatch was the path from Policy to perceived behavioral control, which was significant in the real survey model but not in the LLM-based model. This suggests that, in the survey data, policy arrangements such as clearer governance and ownership support increased residents’ perceived  behavioral control, whereas LLM-based agents did not fully reproduce this mechanism. Instead, policy-related considerations in the LLM-based model more often strengthened supportive attitudes and subjective norms.

In terms of effect-size recovery, the overall magnitude alignment was acceptable, with a mean LLM-to-survey coefficient ratio of 1.11 and a median ratio of 0.94. However, the coefficient-difference tests indicate that this agreement was stronger for signs and significance than for exact effect magnitudes. Nine of the 19 paths showed statistically significant differences between the LLM-estimated and survey-based coefficients. The LLM model tended to overestimate several economic or attitudinal mechanisms while underestimating several social or policy-related mechanisms. These results suggest that the simulated agents reproduced the qualitative structure of the survey-based behavioral model well, but still differed from human respondents in the relative strength assigned to specific mechanisms.
\begin{table}[htbp]
\centering
\caption{Comparison of survey-based effects and LLM-estimated effects.}
\label{tab:real_llm_comparison_grouped}
\small
\setlength{\tabcolsep}{4pt}
\begin{tabular}{llrrrrrr}
\toprule
Outcome & Predictor & Real Est. & Real $p$ & LLM Est. & LLM $p$ & $z_{\mathrm{diff}}$ & $p_{\mathrm{diff}}$ \\
\midrule
\multirow{4}{*}{Attitude}
& Economic & 0.172**  & 0.006 & 0.383*** & \textless 0.001 & 3.61  & \textless 0.001 \\
& Location & 0.210**  & 0.004 & 0.304*** & \textless 0.001 & 1.26  & 0.209 \\
& Social   & 0.290*** & 0.001 & 0.101*** & \textless 0.001 & -2.25 & 0.025 \\
& Policy   & 0.176**  & 0.007 & 0.084*** & \textless 0.001 & -1.67 & 0.096 \\
\midrule
\multirow{4}{*}{PBC}
& Economic & 0.135*   & 0.034 & 0.507*** & \textless 0.001 & 6.35  & \textless 0.001 \\
& Location & 0.198**  & 0.007 & 0.287*** & \textless 0.001 & 1.20  & 0.229 \\
& Social   & 0.221**  & 0.009 & 0.042*   & 0.044          & -2.14 & 0.032 \\
& Policy   & 0.203**  & 0.002 & 0.027    & 0.197          & -3.19 & 0.001 \\
\midrule
\multirow{4}{*}{SubjectiveNorm}
& Economic & 0.190**  & 0.002 & 0.223*** & \textless 0.001 & 0.49  & 0.621 \\
& Location & 0.172*   & 0.016 & 0.366*** & \textless 0.001 & 2.31  & 0.021 \\
& Social   & 0.352*** & 0.001 & 0.331*** & \textless 0.001 & -0.21 & 0.830 \\
& Policy   & 0.147*   & 0.021 & 0.087*** & \textless 0.001 & -0.97 & 0.330 \\
\midrule
\multirow{7}{*}{Agree}
& Economic       & 0.192*** & 0.001 & 0.063*   & 0.026          & -2.06 & 0.040 \\
& Location       & 0.127*   & 0.044 & 0.153*   & 0.011          & 0.28  & 0.779 \\
& Social         & 0.169*   & 0.031 & 0.091*** & \textless 0.001 & -0.86 & 0.388 \\
& Policy         & 0.158**  & 0.006 & 0.047**  & 0.004          & -2.03 & 0.043 \\
& Attitude       & 0.161*   & 0.014 & 0.364*** & \textless 0.001 & 2.58  & 0.010 \\
& SubjectiveNorm & 0.142*   & 0.035 & 0.160*** & \textless 0.001 & 0.26  & 0.792 \\
& PBC            & 0.146*   & 0.011 & 0.059**  & 0.002          & -1.24 & 0.215 \\
\bottomrule
\end{tabular}

\vspace{0.5em}
\begin{minipage}{0.9\linewidth}
 * $p<0.05$, ** $p<0.01$, *** $p<0.001$.
\end{minipage}
\end{table}

\clearpage

\subsection{Application case}

\subsubsection{Study area}
We use Huadu District in Guangzhou, Guangdong Province, as the empirical case to show the application of CoRenew. The study focuses on 14 residential communities identified as potential targets for urban renewal. Full information on the 14 communities and the dataset sources used are listed in Appendix \ref{app:community_info}. We classify communities along three dimensions highlighted in prior research \citep{plaut2010decisions,poteete2004heterogeneity,muczynski2023collective}: average resident annual income, housing appreciation potential, and within-community income heterogeneity. 

%% show some application cases

\subsubsection{Global policy optimization across communities} \label{optimaloverallpolicy}

The platform enables users to search over alternative global policy combinations and identify the optimal rule under multiple objectives across all projects. We define the search space using extension caps of 20\%, 30\%, 40\%, and 50\%, crossed with subsidy caps of 0\%, 5\%, 8\%, 12\%, 16\%, and 20\%. The search proceeds in two stages. Phase 1 screens all combinations on a 20\% sample of communities using two random seeds and a shortened four-round negotiation horizon. Phase 2 then re-evaluates the 12 surviving candidate rules on the full set of 14 communities using five random seeds. Performance is evaluated against five objectives: maximizing negotiation success, minimizing total subsidy cost, maximizing low-income utility, minimizing utility inequality, and maximizing developer profit. Appendix Table \ref{tab:policy_rules_ranked} summarizes the evaluation results for each policy combination. Among them, the best-performing combination is a 30\% extension cap paired with a 12\% subsidy cap, which delivers the strongest overall performance across communities. 

\begin{figure*}[htbp]
    \centering
    \includegraphics[width=1\textwidth]{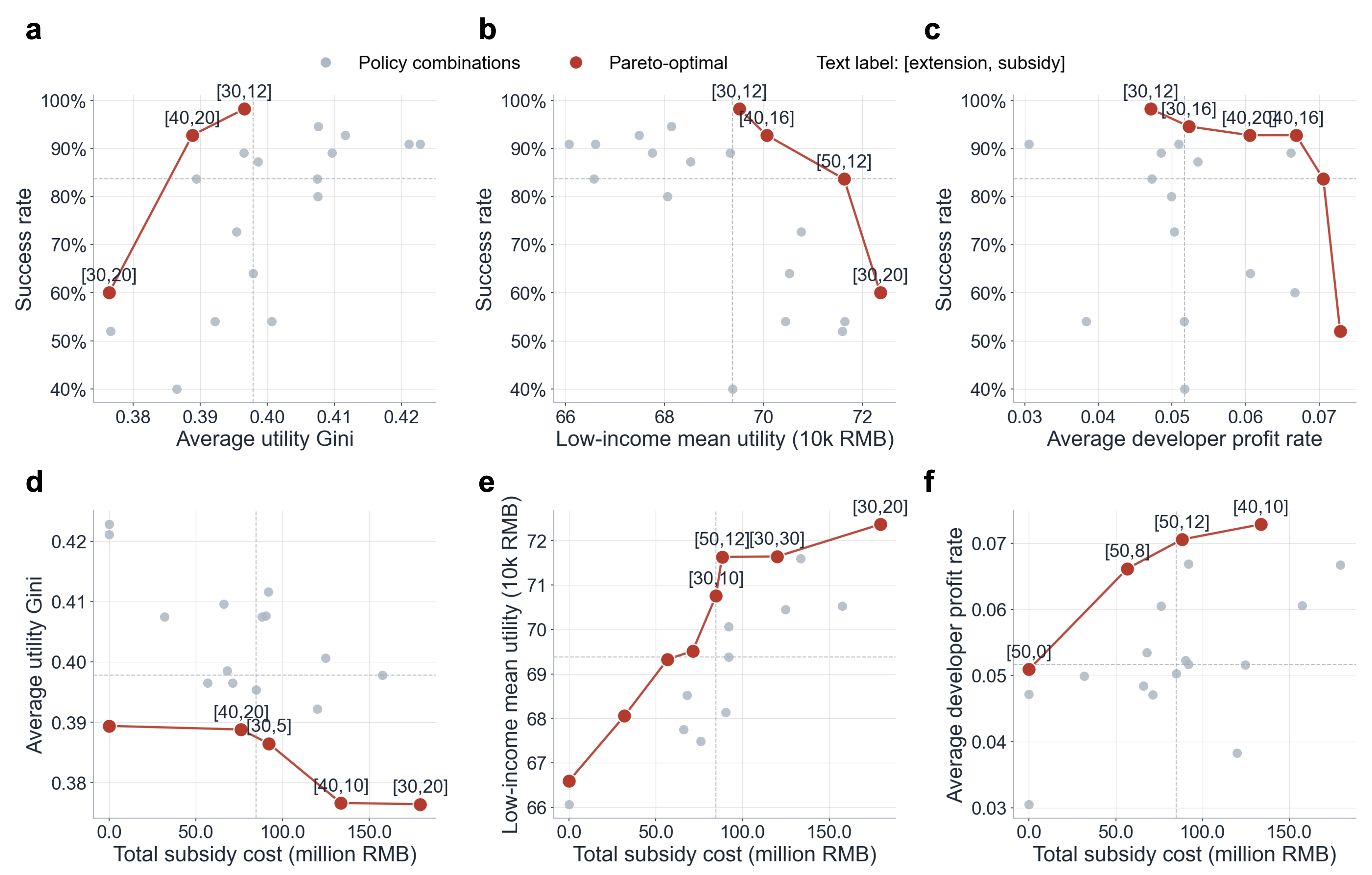}
    \caption{Pareto frontiers of policy combinations under pairwise policy trade-offs.
}
    \label{fig:pareto}
\end{figure*}

Users can also compare a set of policies under two selected objectives. Figure \ref{fig:pareto} summarizes the multi-objective evaluation of policy combinations and reveals a set of clear trade-offs across policy goals. Higher redevelopment success is consistently associated with weaker performance on distributional outcomes. As more communities reach agreement, utility inequality among residents tends to increase and the mean utility of low-income residents tends to decline (Figure \ref{fig:pareto}a-b). The results also suggest a practical profitability boundary for developers (Figure \ref{fig:pareto}c). Higher agreement generally requires developers to accept lower margins and offer more favorable terms to residents, whereas policies associated with higher developer profitability tend to produce lower redevelopment success.

CoRenew also provides insights for public subsidy scheme design. The results show that public subsidy reshapes these trade-offs by improving equity and supporting more vulnerable residents (Figure \ref{fig:pareto}d-e). Higher subsidy levels are associated with lower utility inequality and higher utility for low-income households, but these effects appear to be threshold-dependent. Equalizing effects become more visible only after subsidy reaches a higher range, and the effects on low-income utility show diminishing marginal returns. The objective trade-off analysis also highlights the strategic interaction between planner and developer incentives (Figure \ref{fig:pareto}f). Higher public subsidy is associated with higher developer profit rates, indicating that part of the fiscal transfer is absorbed into the bargaining position of market actors rather than being translated directly into resident welfare.

\subsubsection{Community-specific policy optimization}
\begin{figure*}[htbp]
    \centering
    \includegraphics[width=1\textwidth]{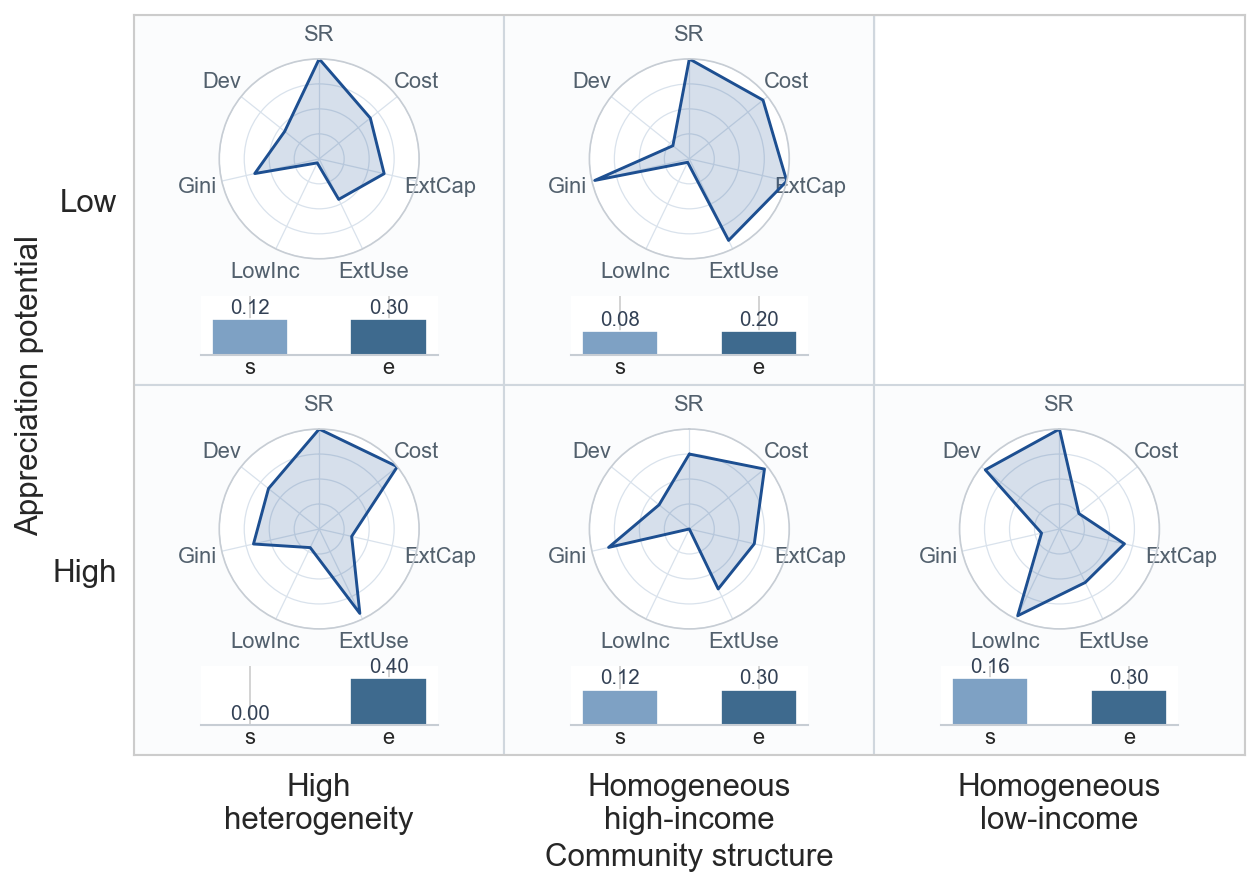}
    \caption{Optimal policy combinations for different community types and their objective-specific performance.}
    \label{fig:optimal_category}
\end{figure*}

CoRenew can also identify local optimal policies for specific communities. Figure \ref{fig:optimal_category} summarizes the optimal policy identified for each community type and the corresponding multi-objective performance. The results show that low-appreciation, high-income homogeneous communities require only a lighter intervention (0.08, 0.20). High-heterogeneity communities are best served by a pure extension strategy (0.00, 0.40), with rising redevelopment gains generating sufficient internal surplus to replace subsidy. High-income homogeneous communities with high appreciation potential, however, revert to a mixed package (0.12, 0.30), because larger expected gains intensify bargaining over surplus. Low-income homogeneous communities move in the opposite direction, requiring the strongest subsidy-supported package (0.16, 0.30). In these communities, appreciation alone is not enough: limited purchasing power keeps affordability constraints binding even under strong redevelopment upside.

\clearpage
\subsubsection{Assessment of semantic policies}

The system allows users to test semantic policy inputs. In this case, we implemented five policy scenarios expressed entirely in natural language: (1) escrow account, which increases transparency in fund management and implementation procedures; (2) owner supervision, which allows residents to participate in construction oversight; (3) shared ownership, which distributes redevelopment costs and future returns between residents and the government according to ownership shares; (4) media publicity, which increases residents’ exposure to redevelopment-related information and public discussion; and (5) moving volunteer service, which provides practical support during relocation and implementation.

\begin{figure*}[htbp]
    \centering
    \includegraphics[width=1\textwidth]{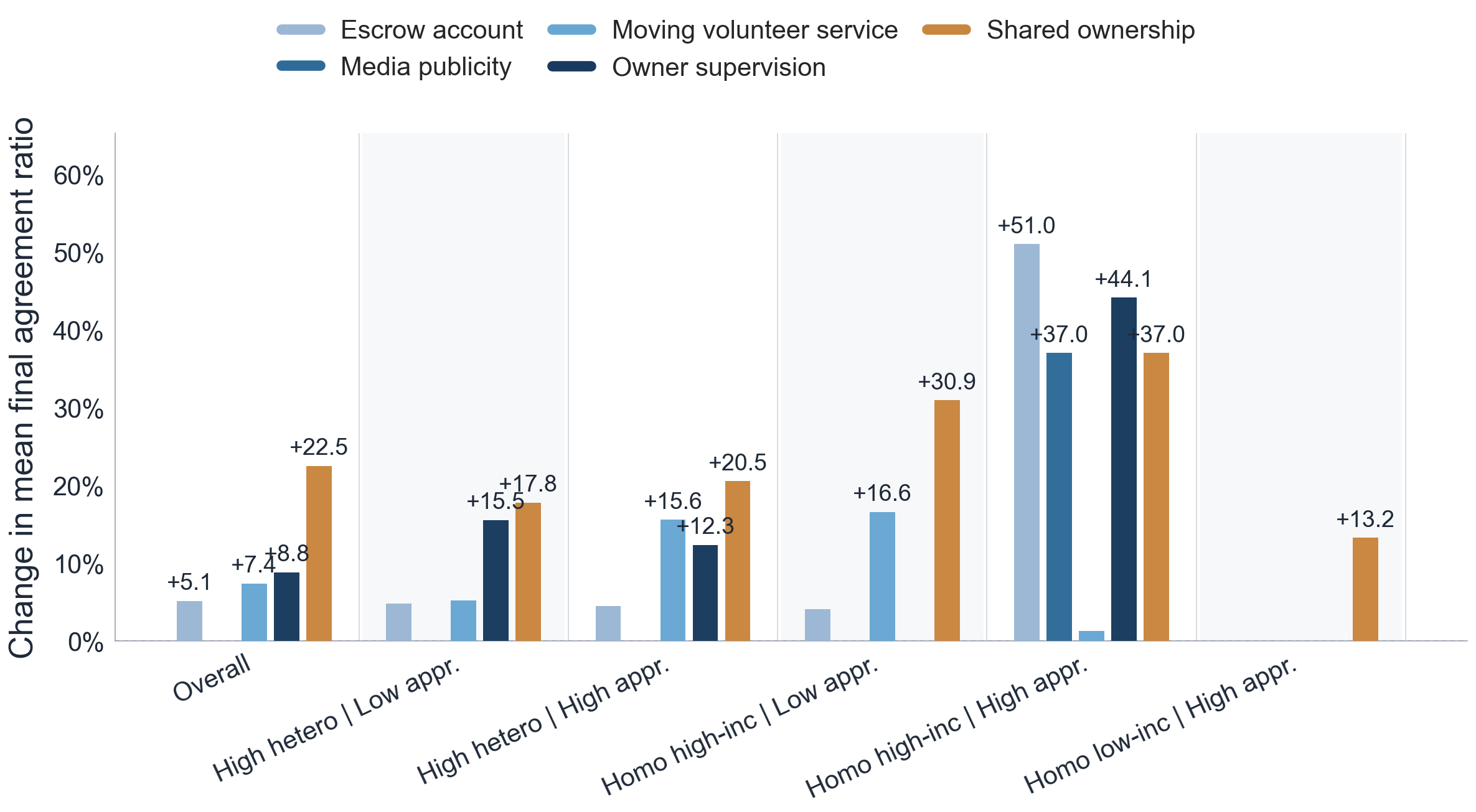}
    \caption{Effects of natural-language policy interventions on agreement outcomes across community types.}
    \label{fig:semantic_policy}
\end{figure*}

All policy interventions increased agreement to some extent, but their effects varied markedly across community types (Figure \ref{fig:semantic_policy}). Shared ownership produced the strongest overall effect, whereas media publicity was consistently the weakest. In low-income homogeneous communities, where affordability is the main barrier to agreement, shared ownership was the only intervention that proved effective. In high-income homogeneous communities with strong appreciation potential, as well as in high-heterogeneity communities, the most effective interventions shifted toward escrow accounts and owner supervision. In these communities, agreement depends less on direct cost sharing than on policies that enhance institutional trust, procedural transparency, and implementation credibility.

\clearpage
\section{Conclusion and Discussion}
%% Conclusion
This study introduced CoRenew, an LLM-based simulation platform for multifamily residential redevelopment policy assessment. Driven by LLM-based agents, CoRenew simulates how residents in different communities respond to policy through multi-round negotiation. By connecting macro-level policy design with micro-level bargaining responses, it helps address both the underrepresentation of negotiation dynamics and the limited treatment of semantic policy space in existing simulation approaches. The platform includes built-in visualization tools and multi-format export functions, enabling direct use in analysis, communication, and policy reporting. We validated the platform against both observed negotiation dynamics and human survey data. CoRenew closely reproduced the empirical negotiation trajectory in a real-world case and showed strong consistency with survey-based path estimates, with accurate recovery of benchmark paths. These results suggest that CoRenew can reasonably reproduce both the process dynamics and behavioral mechanisms of redevelopment negotiation.

Building on this validation, our application in Huadu District, Guangzhou, further demonstrates how CoRenew can support practical policy consultation for multifamily residential redevelopment. CoRenew reproduced mechanisms previously identified in qualitative and policy research, such as the effectiveness of cross-subsidization in high-appreciation settings and the importance of shared-ownership arrangements in low-income communities, where affordability constraints remain central \citep{mukhija2015tradeoffs}. At the same time, the simulations generated new hypotheses about the context-dependent role of heterogeneity: it may facilitate agreement in high-appreciation settings but hinder coordination in homogeneous high-income communities where residents progressively demand larger extension areas \citep{poteete2004heterogeneity,gao2016does}. These findings suggest that CoRenew can extend policy learning beyond single-case evidence by examining qualitative mechanisms and generating new insights across diverse redevelopment contexts. 

%% Flexible
Beyond the present application, CoRenew can also support extended analysis. By retaining full negotiation logs, the platform allows researchers to examine negotiation details between stakeholders, trace how agreement emerges over time, identify pivotal actors, and evaluate effective strategies for different resident groups. It also enables analysis of policy timing and sequencing, making it possible to assess when planner intervention is most effective and which policy instruments work best at different stages of negotiation. In addition, CoRenew is designed to be adaptable across diverse multi-owner housing redevelopment contexts. Key parameters, including cost-benefit structures, resident contribution ratios, developer profit thresholds, and approval rules, can be reconfigured to reflect different legal, financial, and regulatory settings. For example, Singapore’s en bloc sale system, which relies on majority consent rather than direct resident co-financing, can be represented through a high approval threshold and a zero resident contribution ratio \citep{chia2023redevelopment}. In Finland, where added floor area and infill development are often used to finance major repairs in housing companies, redevelopment can be modeled through an extension-based mechanism without subsidy \citep{puustinen2015infill}. In German housing cooperatives, where financing depends more heavily on collective capital and member shares, the platform can instead represent redevelopment through institution-specific arrangements, such as common savings or reserve-fund schemes, without relying on subsidy or extension \citep{balmer2018housing}.

%% Limitations
Several limitations should be acknowledged. First, although CoRenew incorporates TPB-based behavioral mechanisms, the model inevitably simplifies highly contextual negotiation processes. Real-world redevelopment negotiations are shaped by factors such as neighborhood relations, accumulated distrust, emotional responses, risk perceptions, and trust in public and market actors. Therefore, CoRenew should be understood as an exploratory model for policy comparison rather than a precise predictor of negotiation outcomes. Second, model validation is still constrained by the limited availability of detailed negotiation data, as round-by-round records of resident negotiations remain rare in emerging redevelopment practices. Future research should incorporate more cases across different institutional and cultural contexts to test the robustness and transferability of the framework. Third, limitations inherent to LLM-based agents must be recognized. Model outputs may vary with model versions, prompt design, temperature settings, memory mechanisms, and context length. Simulated agents may also reproduce stereotypes or provide superficial representations of behavior \citep{hu2025generative}. Finally, the current version represents interaction mainly through aggregate agreement rates and structured numerical outputs. This improves efficiency for large-scale policy evaluation but reduces behavioral realism by limiting verbal persuasion, informal communication, and social-network effects. Future versions of CoRenew will incorporate richer natural-language interactions and explicit social network structures to better capture how influence, trust, and information diffusion shape negotiation dynamics.
\bibliographystyle{elsarticle-harv}
\bibliography{cas-refs}

\end{document}